\documentclass[condensedmatter,article,submit,moreauthors,pdftex,10pt,a4paper]{mdpi} 
\usepackage{graphicx}
\usepackage{float}
\usepackage{latexsym}
\usepackage{amssymb}
\usepackage{amsfonts}
\usepackage{amsmath}
\usepackage{amssymb}

\firstpage{1} 
\articlenumber{x}
\doinum{10.3390/------}
\pubvolume{xx}
\pubyear{2017}
\copyrightyear{2017}
\history{Received:}

\Title{Inverse spin galvanic effect in the presence of impurity spin-orbit scattering: a diagrammatic approach}

\Author{Amin Maleki and Roberto Raimondi$^{\star}$}

\AuthorNames{Amin Maleki, and Roberto Raimondi}

\address{%
 \quad Dipartimento di Matematica e Fisica, Universit\`a Roma Tre, Via della Vasca Navale 84, 00146 Roma, Italy.}

\corres{Correspondence: email: roberto.raimondi@uniroma3.it; Tel. +39 06 5733 7032}

\abstract{Spin-charge interconversion is currently the focus of intensive experimental and theoretical research both for its intrinsic interest and for its potential exploitation in the realization of new spintronic functionalities. Spin-orbit coupling is one of the key microscopic mechanisms to couple charge currents and spin polarizations.
The Rashba spin-orbit coupling in a two-dimensional electron gas has been shown to give rise to the inverse spin galvanic effect, i.e. the generation of  a non-equilibrium spin polarization by a charge current. Whereas the Rashba model may be applied to the interpretation of experimental results in many cases, in general in a given real physical system spin-orbit coupling also occurs due other mechanisms  such as Dresselhaus bulk inversion asymmetry and scattering from impurities.
In this work we consider the inverse spin galvanic effect in the presence of Rashba, Dresselhaus and impurity spin-orbit scattering. We find that the size and form of the inverse spin galvanic effect is greatly modified by the presence of the various sources of spin-orbit coupling. 
Indeed, spin-orbit coupling affects  the spin relaxation time by adding the Elliott-Yafet mechanism to the Dyakonov-Perel and, furthermore, it changes the
non-equilibrium value of the current-induced spin polarization by introducing a new spin generation torque.
We use a diagrammatic Kubo formula approach to evaluate the spin polarization-charge current response function.  
We finally comment about the relevance of our results for the interpretation of experimental results.}


\def\be{\begin{equation}}
\def\ee{\end{equation}}
\def\ber{\begin{eqnarray}}
\def\eer{\end{eqnarray}}

\def\sigmabold{\mbox{\boldmath $\sigma$}}
\def\rv{{\bf r}}

\def\pv{{\bf p}}

\def\br{{\bf r}}

\def\bk{{\bf k}}

\def\nn{\nonumber}
\def\nn{\nonumber}

\newcommand{\commentout}[1]{}

\DeclareMathAlphabet\mathbfcal{OMS}{cmsy}{b}{n}

\def\be{\begin{equation}}
\def\ee{\end{equation}}
\def\ber{\begin{eqnarray}}
\def\eer{\end{eqnarray}}
\def\nn{\nonumber}
\def\bk{{\bf k}}

\begin{document}
\section{Introduction}
\label{sec_intro}
 The spin galvanic effect  and its  inverse manifestation  have been intensively investigated over the past decade both for their intrinsic fundamental interest \cite{Ganichev2016} and for their application potential in future generation electronic and spintronics technology \cite{Ando2017,Soumyanarayanan2016}. 
The non-equilibrium generation of a spin polarization perpendicular to  an externally applied electric field  is referred to as the  inverse spin galvanic effect (ISGE), whereas the spin galvanic effect (SGE) is its Onsager reciprocal, whereby  a spin polarization injected through a nonmagnetic
material creates a charge current in the  direction perpendicular to the spin polarization. As an all-electrical method of generating and detecting spin polarization in nonmagnetic materials, both these effects may be used for applications such as spin-based field effect transistors \cite{Gardelis1999,DasSarma2001207,Sugahara2005,Koo1515} and magnetic random access memory (MRAM) \cite{Miyazaki1995,Yuasa2004}.

The ISGE, also known as Edelstein effect or current-induced spin polarization, was originally proposed by Ivchenko and Pikus \cite{ivchenko1978new}, and observed by Vorob'ev et al. in tellurium \cite{1979JETPL..29..441V}. Later the ISGE was theoretically analyzed by Edelstein in a two-dimensional electron gas (2DEG) with Rashba spin-orbit coupling (SOC) \cite{edelstein1990solid} and also by Lyanda-Geller and Aronov\cite{aronov1989nuclear}. Notice that the SGE in the spin-charge conversion is sometimes referred to as the inverse Rashba-Edelstein effect. 
The SGE has been observed experimentally in GaAs QWs by Ganichev et al. \cite{RIS_0Spin-galvanic}, where the spin polarization was detected by measuring the current produced by circularly polarized light. 
In semiconducting structures the ISGE can be measured by optical methods such as Faraday rotation, linear-circular dichroism in transmission of terahertz radiation and time resolved Kerr rotation \cite{PhysRevLett.86.4358,Ganichev2006127,PhysRevLett.96.186605,Ganichev2016}. Very recently, a new way of converting spin to charge current has been experimentally developed by Rojas-S\'anchez et al., where, by the spin-pumping  technique, the non-equilibrium spin polarization injected from a ferromagnet   into a silver (Ag)/Bismuth (Bi) interface yields an electrical current \cite{Sanchez13}. Successively, the SGE has also been observed in many interfaces with strong spin-orbit splitting, including metals with semiconductor giant SOC or insulator such as Fe/GaAs \cite{chen2016}, \textnormal{$Cu/Bi_2O_3$}\cite{1882-0786-9-3-033001}.

Generally speaking, the SGE can be understood phenomenologically by symmetry arguments. Electrical currents and spin polarizations are polar and axial vectors, respectively. In centro-symmetric systems, polar and axial vectors transform differently and no SGE effect is expected. In restricted symmetry conditions, however,
polar and axial vectors components may transform similarly. 
Consider, for instance, the case of electrons confined in the xy plane with 
 the mirror reflection through  the yz plane. Under such a symmetry operation, the electrical currents along the the x and y directions transform
  as $J_x \rightarrow -J_x$ and $J_y \rightarrow J_y$. The spin polarizations transform as the components of angular momentum, and we have
  $S^y \rightarrow -S^y$ and $S^x\rightarrow S^x$. Hence, one expects a coupling between $J_x$ and $S^y$ or between $J_y$ and $S^x$.
Such a coupling is the SGE.

At microscopic level the strength of the coupling is due to the SOC.
 Usually the SOC is classified as extrinsic and intrinsic, depending on the origin of the electrical potential. The intrinsic SOC arises due to the crystalline potential 
 of the host material or due the confinement potential  associated with the device structure.  On the other hand,  the extrinsic SOC is due to the atomic potential of random impurities, which determine the transport properties of a given material.  
The majority of the studies on SGE/ISGE has focused on the Rashba SOC (RSOC) for electrons moving in the xy plane, which was  originally introduced by Rashba \cite{Rashba1960} to study the properties of the energy spectrum of non-centrosymmetric crystals of the CdS type and later successfully applied to the interpretation  the two-fold spin splitting of electrons and holes in asymmetric semiconducting heterostructures \cite{bychkov1984properties}. RSOC is classified as due to structure inversion asymmetry (SIA), which is responsible for the confinement of electrons in the xy plane. In addition one may also consider the SOC arising from the bulk inversion asymmetry (BIA), usually referred to as Dresselhaus SOC (DSOC) \cite{Winkler2003}. 
Both RSOC and DSOC modify the energy spectrum by introducing a momentum-dependent spin splitting. This also can be understood quite generally on the basis of symmetry considerations. In a solid spin degneracy for a couple of states with opposite spin and with cristalline wave vector $\bk$ is the result of both time reversal invariance and parity (space inversion invariance). By breaking the parity, as for instance, in a confined two-dimensional electron gas,  the spin degeneracy is lifted and
the Hamiltonian acquires an effective momentum-dependent magnetic field, which is the SOC. As a result electron states can be classified with their chirality in the sense that their   spin state depends on their wave vector.  In a such a situation, scalar disorder, although not directly acting on the spin state, influences the spin dynamics by affecting the wave vector of the electrons and holes. Spin relaxation arising in this context is usually referred to as the Dyakonov-Perel (DP) mechanism.

Extrinsic SOC  originates from the potential which is responsible for the scattering from an impurity. In this case, before and after the scattering event, there is no direct connection 
between the wave vector and the spin of the electron.  The scattering amplitude can be divided in  spin-independent and spin-dependent contributions
\be
S_{\pv,\pv'}= A+{\hat \pv}\times {\hat \pv'}\cdot {\boldsymbol\sigma} B\label{scatteringamplitude},
\ee
where $\hat \pv$ and $\hat \pv'$ are the unit vector along the direction of the momentum before and after the scattering and ${\boldsymbol\sigma}$ is the vector of the Pauli matrices. As explained in Ref.\cite{Lifshits2009}, different combinations of the amplitudes $A$ and $B$ correspond to specific physical processes.
The $|A|^2+|B|^2$ describes the total scattering rate, whereas $|B|^2$ is associated to the Elliott-Yafet (EY) spin relaxation rate. Interference terms between the two amplitudes yield coupling among the currents. More in detail, the combination $A B^*+A^* B$ describes the skew scattering, which is responsible for the coupling between the charge and spin currents, whereas $A B^*-AB^*$ gives rise to the swapping of spin currents.

As noted in Ref.\cite{Ganichev2016},  when both intrinsic and extrinsic SOC is present, the non-equilibrium spin polarization of the ISGE
depends on the ratio of the DP and EY spin relaxation rates.  This was analyzed in Ref.\cite{Raimondi2012} by means of the  Keldysh non-equilibrium Green function within a SU(2) gauge theory-description of the SOC.  Successively, a parallel analysis by standard Feynman diagrams for the Kubo formula was carried out in
Ref.\cite{maleki2016edelstein}. These theoretical studies indeed confirmed that the ratio of DP to EY spin relaxation is able to tune the value of the ISGE.
Such tuning is also affected by the value of the spin Hall angle due to the fact that  spin polarization and spin current are coupled in the presence of intrinsic RSOC.

Recently,  it has been shown  theoretically \cite{gorini2017theory} that the interplay of intrinsic and extrinsic SOC gives rise to an additional spin torque in the Bloch equations for  the spin dynamics and affects the value of the ISGE. This additional spin torque, which  is proportional to both the EY spin relaxation rate and to the coupling constant of RSOC, in Ref.\cite{gorini2017theory}  has been derived in the context of the SU(2) gauge theory formulation mentioned above.
Although the SU(2) gauge theory is a very powerful approach, 
in order to emphasize the physical origin of this new torque 
 it is very useful to show also how the same result can be obtained independently  by using the diagrammatic approach of the Kubo linear response theory. This is the aim of the present paper. 

In this paper we obtain an analytical formula of the ISGE in the presence of the Rashba, Dresselhaus and impurity SOC. In a 2DEG we will show that the intrinsic and extrinsic SOC act in parallel as far as relaxation to the equilibrium state is concerned.

The model Hamiltonian for a 2DEG in the presence of SOC reads
\begin{eqnarray}
H=\frac{p^2}{2m}+\alpha (p_y\sigma_x-p_x\sigma_y)+\beta (p_x\sigma_x- p_y\sigma_y) + V(\rv)-\frac{\lambda_0^2}{4} \nabla V(\rv)\times \pv\cdot \sigmabold ,\label{ham}
\end{eqnarray}
where $\pv=(p_x,p_y)$ is the vector of the components of the momentum operator, $\sigmabold=(\sigma_x,\sigma_y,\sigma_z)$ and $\br$ are the Pauli matrices and the coordinate operators. $m$ is the effective mass, $\alpha$ and $\beta$ are the Rashba and Dresselhaus SOC constants. $V(\br)$ represents a short-range impurity potential and finally $\lambda_0$ is the effective Compton wave length describing the strength of the extrinsic SOC.  We assume the standard model of white-noise disorder potential with $\langle V(\br) \rangle=0$ and Gaussian  distribution  given by $\langle V(\rv)V(\rv')\rangle=n_i v_0^2\delta(\rv-\rv')= (\hbar /(2\pi N_0 \tau_0) )\delta(\rv-\rv')$.  $N_0=m/2\hbar^2 \pi$, $n_i$  and  $v_0$  are the single-particle density of states per spin  in the absence of  SOC, the impurity concentration  and  the scattering amplitude, respectively. 
$\tau_0$ is the elastic scattering time at the level of the Fermi Golden Rule.
From now on  we work with units such that $\hbar=1$.  

The layout of the paper is as follows. In the next Section we formulate the  ISGE (the SGE can be obtained similarly by using the Onsager relations) in terms of the Kubo  linear response theory. In Section \ref{Rasha_model} we derive an expression for the ISGE in the presence of the RSOC and extrinsic SOC. 
This case with no DSOC, whereas it is important by itself, allows to understand the origin of the additional spin torque in a situation which technically simpler to treat with respect to the general case when both RSOC and DSOC are different from zero.
In Section \ref{R/D_model}, we expand our result to the specific case when the both RSOC and DSOC, as well as SOC from impurities, are present. We show how our result can be seen as the stationary solution of the Bloch equations for the spin dynamics. We comment briefly on the relevance of our result for the interpretation of the experiments. Finally, we state our conclusions in Section \ref{conclusions}.

\section{Linear response theory}\label{sec_LRT}

In this Section we use the standard Kubo formula of linear response theory to derive the ISGE in the presence of extrinsic and intrinsic SOC. 
The in-plane spin polarization to linear order in the electric fields is given by
\begin{equation}
	\label{LRT}
	S^i=\sigma^{ij}_{EC}E_j, \  i,j=x,y,
\end{equation}
where $E_i$ is the external electric fields with frequency $\omega$ and $\sigma^{ij}_{EC}$ is the frequency-dependent "Edelstein conductivity"\cite{EDELSTEIN1990233} given by the Kubo formula \cite{shen2014microscopic}
\begin{equation}
	\sigma^{ij}_{EC}(\omega)=\frac{(-e)}{2\pi} \sum\limits_{\pv}  {\rm Tr} [G^A(\epsilon+\omega)\Upsilon_i (\epsilon, \omega) )G^R(\epsilon)J_j ],
	\label{Kubo}
\end{equation}
where the trace symbol includes the summation over spin indices. 
We keep the frequency dependence of  $\sigma^{ij}_{EC}(\omega)$ in order to obtain the Bloch equations for the spin dynamics.
 In Eq.(\ref{Kubo}), $\Upsilon_i (\epsilon, \omega)$ is the renormalized spin vertex relative to a polarization along the $i$ axis, required by the standard series of ladder diagrams of the impurity technique \cite{Schwab2002,Raimondi2005}. 
 $J_j$ are the {\it bare} number current vertices. In the plane-wave basis their matrix element from state $\pv'$ to state $\pv$ read
\begin{eqnarray}
	J_{x}&=&\delta_{\pv,\pv'}\left( \frac{p_x}{m}-\alpha \sigma_y+\beta \sigma_x\right)+\delta j_{x,\pv\pv'},\label{num_curr_x}\\
	J_{y}&=&\delta_{\pv,\pv'}\left(\frac{p_y}{m}+\alpha \sigma_x-\beta \sigma_y\right) +\delta j_{y,\pv\pv'}.\label{num_curr_y}
\end{eqnarray}      
The latter term $\delta J_{j,\pv\pv'}$ in Eqs.(\ref{num_curr_x}-\ref{num_curr_y}), which depends explicitly on disorder,  is of order $\lambda_0^2$ and originates from the last term in the Hamiltonian of Eq.(\ref{ham}).  Such a term   gives rise to the side-jump contribution to the spin Hall effect \cite{Engel2005,Tse2006}
due to the extrinsic SOC.  The side-jump and skew-scattering contributions to the spin Hall effect in the presence of RSOC  have been considered in Ref.\cite{Raimondi2009,Raimondi2010952,Raimondi2012}.  A similar analysis of the side-jump and skew-scattering contributions to  the ISGE has been carried out within the SU(2) gauge theory formualtion in Ref.\cite{Raimondi2012} and, more recently, in Ref.\cite{maleki2016edelstein} by standard Kubo formula diagrammatic methods. 
For this reason we will not repeat such an analysis here, where instead we concentrate on the contributions generated by 
  the first term on the right hand side of Eq.(\ref{num_curr_x}-\ref{num_curr_y}).
  
Within the self-consistent Born approximation, the last two terms of the Hamiltonian (\ref{ham})  yield an effective  the self-energy
when  averaging over disorder. The self-energy is diagonal in momentum space and  has two contributions due to the spin independent and spin dependent scattering \cite{EDELSTEIN1990233,PhysRevB.90.245302} 
\begin{eqnarray}
	\label{born}
	\Sigma^R_{tot}(\pv)&\equiv &\Sigma^R_0(\pv)+\Sigma^R_{EY}(\pv)\nn\\ &=& n_iv_0^2\sum_{\pv'}G^R_{\pv'}+n_iv_0^2\frac{\lambda_0^4}{16}\sum_{\pv'}\sigma_zG_{\pv'}^R\sigma_z (\pv\times\pv')^2_z.\label{selfenergy}
\end{eqnarray}
 
Whereas the imaginary part of the first term gives rise to the standard elastic scattering time 
\be
{\it Im}\Sigma^R_0(\pv)=-i 2\pi N_0 n_i v_0^2 =-\frac{i}{2\tau_0},\label{elastictime}
\ee
The second one is responsible for the EY spin relaxation.
From the point of view of the scattering matrix introduced in the previous Section (cf. Eq.(\ref{scatteringamplitude})), the two self-energies contributions
correspond to the Born approximation for the $|A|^2$ and $|B|^2$, respectively.
Given the self-energy (\ref{selfenergy}),   the retarded  Green function is also diagonal in momentum space and can be expanded in the Pauli matrix basis in the form 
\begin{equation}
	G_{\pv}^{R}=G^{R}_0\sigma_0 +G^{R}_x\sigma_x+G^{R}_y\sigma_y,
	\label{greens function}
\end{equation}
where 
\begin{eqnarray}
	G^{R}_0&=&\frac{G^{R}_++G^{R}_-}{2} \nonumber\\
	G^{R}_x&=&(\alpha\hat{p}_y +\beta \hat{p}_x)\frac{G^{R}_+-G^{R}_-}{2\gamma}\nonumber\\
	G^{R}_y&=&-(\alpha \hat{p}_x +\beta \hat{p}_y)\frac{G^{R}_+-G^{R}_-}{2\gamma}.
\end{eqnarray}

In the above  $G^R_{\pm }(\epsilon)=(\epsilon-\frac{p^2}{2m} \mp \gamma p+\frac{i}{2\tau_\pm})^{-1}$ is the Green function corresponding to the two branches in which the energy spectrum  splits due to the SOC.  The factor $\gamma^2=\alpha^2+\beta^2+2\alpha \beta \sin (2\phi)$ with $\hat p_x =\cos (\phi )$ and $\hat p_y= \sin (\phi)$ describes the dependence in momentum space of the SOC, when both RSOC and DSOC are present. Notice that inversion in the two-dimensional momentum space ($(p_x,p_y)\rightarrow (-p_x,-p_y)$) leaves the factor $\gamma$ invariant, since it corresponds to $\phi \rightarrow \phi +\pi$. As a consequence,
$G_{x,y}\rightarrow -G_{x,y}$, whereas $G_0$ is invariant. This observation will turn out to be useful later when evaluating the renormalization of the spin vertices.
The advanced Green function is easily obtained via the relation  $G^A_{\pm }=(G^R_{\pm })^*$.
In the expression for $G^R_{\pm }$, $\frac{1}{2\tau_\pm}$ is a band-dependent time relaxation and plays an important role in our analysis.
In order to obtain this term we note  that, after momentum integration over $\pv'$ in Eq.(\ref{selfenergy}),  the imaginary part of the retarded self-energy reads
\begin{equation}
	\Sigma^R_{\pm}=-{\rm i}\frac{1}{2\tau_0}-{\rm i}\left(\frac{\lambda_0^2}{4}\right)^2\frac{1}{4\tau_0}p_F^2p^2_\pm\equiv -\frac{{\rm i}}{2\tau_{\pm}}
	\label{selfenergy_2}
\end{equation} 
Above,  we indicate with  $p_F$  the Fermi momentum  without RSOC and DSOC and with $p_\pm$  the $\gamma$-dependent momenta of the two spin-orbit split Fermi surfaces. To lowest order in the spin-orbit splitting we have
\begin{equation}
	p_\pm=p_F(1\mp\frac{\gamma}{v_F}),
\end{equation} 
where $v_F=p_F/m$. The momentum factors originate from the square of the vector product  in the second term of Eq.(\ref{selfenergy}).
The factor $p_F^2$ is due to the {\it inner} $\pv'$ momentum, which upon integration is eventually fixed  at the Fermi surface in the absence of RSOC and DSOC. More precisely, when evaluating the momentum integral, one ends up by summing the contributions of the two spin-orbit split bands in such a way that the $\alpha$- and $\beta$-dependent shift of the two Fermi surfaces cancels in the sum. However, the {\it outer} $\pv$ momentum remains unfixed. Its value will be fixed by the poles of the Green function in a successive integration over the momentum. Then, the $\gamma$-dependent relaxation times of the two Fermi surfaces read
\begin{equation}
	\frac{1}{\tau_{\pm}}=\frac{1}{\tau}(1\mp\frac{\tau}{\tau_{EY}}\frac{\gamma}{v_F})
	\label{tau+-},
\end{equation}
where 
\begin{equation}
	\frac{1}{\tau}=\frac{1}{\tau_0}+\frac{1}{2\tau_{EY}},
\end{equation}
with the standard expression for the EY spin relaxation rates 
\begin{equation}
	  \frac{1}{\tau_{EY}}=\frac{1}{\tau_0}\left(\frac{\lambda_0 p_F}{2}\right)^4.
\end{equation}

In order to evaluate Eq.(\ref{Kubo}), we need the renormalized spin vertex  $\Upsilon_i$ which has an expansion in Pauli matrices
$\Upsilon_i=\sum_{\rho=0,1,2,3}\Upsilon_i^{\rho}\sigma_{\rho}$, with 
the {\it bare} spin vertices  $\Upsilon^{(0)}_i=\sigma_i$.  We have dropped the explicit dependence $\Upsilon_i (\epsilon, \omega)$ for simplicity's sake.
For vanishing RSOC or DSOC, symmetry tells that the renormalized spin vertices share the same matrix structure of the bare ones $\Upsilon_i\sim \sigma_i$. However,
when both RSOC and DSOC are present, symmetry arguments again indicate that $\Upsilon_x$ and $\Upsilon_y$ are not simply proportional  to $\sigma_x$ and $\sigma_y$, but acquire    both $\sigma_x$ and $\sigma_y$ components.
By following the standard procedure \cite{PhysRevB.90.245302}, after projecting over the Pauli matrix components, the  vertex equation reads
\begin{eqnarray}
	\Upsilon_i^{\rho}=\delta_{\rho i}+\frac{1}{2}\sum_{\mu \upsilon \lambda} I_{\mu \upsilon} {\rm Tr}[\sigma_\rho \sigma_\mu\sigma_\lambda\sigma_\upsilon]\Upsilon_i^{\lambda}+\frac{1}{2}\sum_{\mu \upsilon\lambda} J_{\mu \upsilon} {\rm Tr}[\sigma_\rho \sigma_z \sigma_\mu\sigma_\lambda\sigma_\upsilon \sigma_z]\Upsilon_i^{\lambda},
	\label{vertex}
\end{eqnarray}  
where 
\begin{equation}
	I_{\mu \upsilon}=\frac{1}{2\pi N_0 \tau_0}\sum_{\pv'} G^A_\mu(\epsilon+\omega) G^R_\upsilon(\epsilon), \    
	J_{\mu \upsilon}=\frac{\tau_0}{2\tau_{EY}}I_{\mu \upsilon}.\label{integrals}
\end{equation}

Once the spin vertices are known, the "Edelstein conductivities" from Eq.(\ref{Kubo}) can be put in the form 
\begin{equation}
	\sigma^{ij}_{EC}=\Upsilon_i^{\rho}\Pi_{\rho j}
	\label{gamma_x1}
\end{equation}
with the {\it bare} "Edelstein conductivities" given by
\begin{equation}
	\Pi_{\rho j}  =\frac{(-e)}{2\pi} \sum_{\pv} {\rm Tr} [G^A(\epsilon+\omega)\frac{\sigma_{\rho}}{2}G^R(\epsilon) J_j]
	\label{gamma y2}.
\end{equation}
The bare  "Edelstein conductivities" are those one would obtain by neglecting the vertex corrections due to the ladder diagrams.
It is useful to point  that one could have adopted the alternative route to renormalize the number current vertices and use the bare spin vertices.
Indeed, this was the route followed originally by Edelstein \cite{EDELSTEIN1990233}. Since, the renormalized  number current vertices, in the DC zero-frequency limit, 
vanish \cite{Raimondi2005}, the evaluation of the Edelstein conductivity reduces to a  bubble with bare spin vertices and the current vertices in absence
of RSOC and DSOC.

\section{Inverse spin-galvanic effect in the Rashba model}
\label{Rasha_model}

To keep the discussion as simple as possible, in this Section we confine first to the case when only RSOC is present. We will derive the spin polarization, $S^y$, when an external electric field is applied along the $x$ direction. Then in the next Section we will evaluate the  Bloch equation in the more general case when both RSOC and DSOC  are present. In the case $\beta=0$, the renormalized spin vertex $\Upsilon^y$ is simply proportional to $\sigma_y$, which means $\Upsilon^y=\Upsilon^y_y\sigma^y$. Upon the integration over momentum in Eq.(\ref{vertex}), only $I_{00}$ is non-zero and other eight possibilities of $(\mu,\nu)$ in $I_{\mu,\nu}$ are zero. The cases   $(0,x/y)$, $(x/y,0)$, $(x,y)$ and $(y,x)$ vanish because of angle integration,  whereas the two other cases $(x,x)$ and $(y,y)$ cancel each other after taking the trace in Eq.(\ref{vertex}).

As a result we finally obtain (in the diffusive approximation $\omega\tau\ll1$)
\begin{eqnarray}
\Upsilon_y=\Upsilon_y^y\sigma^y= \frac{1}{1-I_{00}+J_{00}}\sigma^y=\frac{1-4i\omega\tau}{\frac{\tau}{\tau_{s}}-i\omega\tau} \sigma^y\label{vertexa}
\end{eqnarray}
where the integral $I_{00}$ has been evaluated in the appendix \ref{app}
\begin{eqnarray}
I_{00}=\left(\frac{1-3i\omega\tau-\frac{\tau}{\tau_{\alpha}}}{1-4i\omega\tau}\right)\left(\frac{\tau}{\tau_0}\right)
\end{eqnarray}
with the total spin relaxation rate being $\frac{1}{\tau_s}=\frac{1}{\tau_{EY}}+\frac{1}{\tau_{\alpha}}$.
Here $1/\tau_{\alpha}=(2m\alpha)^2 D$ defines the DP spin relaxation rate due to the RSOC.  
Notice that, in the absence of SOC the vertex becomes singular by sending  to zero the frequency, signaling the spin conservation in that limit.
One sees that the EY and DP relaxation rates simply add up.
 This gives then $\sigma^{yx}=\Upsilon_y^y\Pi_{yx}$. Physically, in the zero-frequency limit, the factor $\Upsilon_y^y=\tau_s /\tau$ counts how many impurity scattering events are necessary to relax the spin.  In the diffusive regime $\tau_s\gg \tau$, i.e. many impurity scattering events are necessary to erase the memory of the initial spin direction.
 
 By neglecting the contribution from the extrinsic SOC in the expression (\ref{num_curr_x}) for the current vertex, the bare conductivity $\Pi_{yx}$ naturally separates in two terms $\Pi^{(A)}_{yx}$ and $\Pi^{(B)}_{yx}$ due to the components $p_x/m$ and 
$-\alpha \sigma^y$ of the number current vertex.
The expression for $\Pi^{(A)}_{yx}$ reads
\begin{eqnarray}
	\Pi^{(A)}_{yx}&=& (-e)\frac{1}{2\pi}\sum_{\bf p} {\rm Tr} \left[ G^A(\epsilon+\omega) \frac{\sigma^y}{2}G^R(\epsilon) \frac{p_x}{m}\right]\nonumber \\
	&=&\frac{e}{4\pi m}\sum_{\bf p}\frac{p}{2}\left[ G^A_{+}(\epsilon+\omega) G^R_{+}(\epsilon)-G^A_{-}(\epsilon+\omega)G^R_{-}(\epsilon)\right]\nonumber\\
	&=& \frac{e}{4m}\left(\frac{p_+N_+}{-i\omega+\frac{1}{\tau_+}}-\frac{p_-N_-}{-i\omega+\frac{1}{\tau_-}}\right).
\end{eqnarray}
In the above $p_{\pm}$, $N_{\pm}$ and $\tau_{\pm}$ refer to the Fermi momentum, density of states and quasiparticle time in the $\pm$-band. To order $\alpha /v_F$, one has 
\begin{equation}
	\label{bands_parameters}
	p_{\pm}=p_F (1\mp \alpha /v_F), \ N_{\pm}=N_0 (1\mp \alpha /v_F).
\end{equation}
By including the contribution of the quasiparticle time in the $\pm$-band from Eq.(\ref{tau+-}), one gets
\begin{equation}
\label{ed_a_2}
\Pi^{(A)}_{yx}= S_0 \left( \frac{1-\frac{\tau }{2\tau_{EY}}-i\omega\tau}{1-2i\omega\tau}\right),
\end{equation}
where $S_0=-e N_0 \alpha \tau$.
The evaluation of  $\Pi^{(B)}_{yx}$ is more direct. It gives
\begin{eqnarray}
	\Pi^{(B)}_{yx}&=& \frac{e\alpha}{2\pi}\sum_{\bf p} {\rm Tr} \left[ G^A(\epsilon+\omega) \frac{\sigma^y}{2}G^R(\epsilon)\sigma^y\right]\nonumber \\
	&=& \frac{e\alpha}{2\pi}\sum_{\bf p}\left( G^A_{0}(\epsilon+\omega) G^R_{0}(\epsilon)\right)\nonumber\\
	&=&-S_{0}\left( \frac{1-\frac{\tau }{\tau_{\alpha}}-3i\omega\tau}{1-4i\omega\tau}\right)
\end{eqnarray}
 Combining both contributions with accuracy up to order $\omega\tau$ gives
 \begin{eqnarray}
\Pi_{yx}=\Pi^{(A)}_{yx}+\Pi^{(B)}_{yx}=S_0 \left(\frac{ \frac{\tau
	}{\tau_{\alpha}}-\frac{\tau}{2\tau_{EY}}}{1-6i\omega\tau}\right)
 \end{eqnarray}

By combining  the vertex correction Eq.(\ref{vertexa}) and the bare conductivity $\Pi_{yx}$ in Eq.(\ref{gamma_x1}),  we get following contribution to the frequency-dependent spin polarization
\begin{equation}
	\label{ed_self}
(S^y)^{(1)}=\frac{1}{(\frac{\tau}{\tau_s}-i\omega\tau)} \left(\frac{1-4i\omega\tau}{1-6i\omega\tau}\right)S_{\alpha}^x  \left( \frac{\tau
	}{\tau_{\alpha}}-\frac{\tau}{2\tau_{EY}}\right).
\end{equation}

\begin{figure}[t]
	\begin{center}
		\includegraphics[width=1.5in]{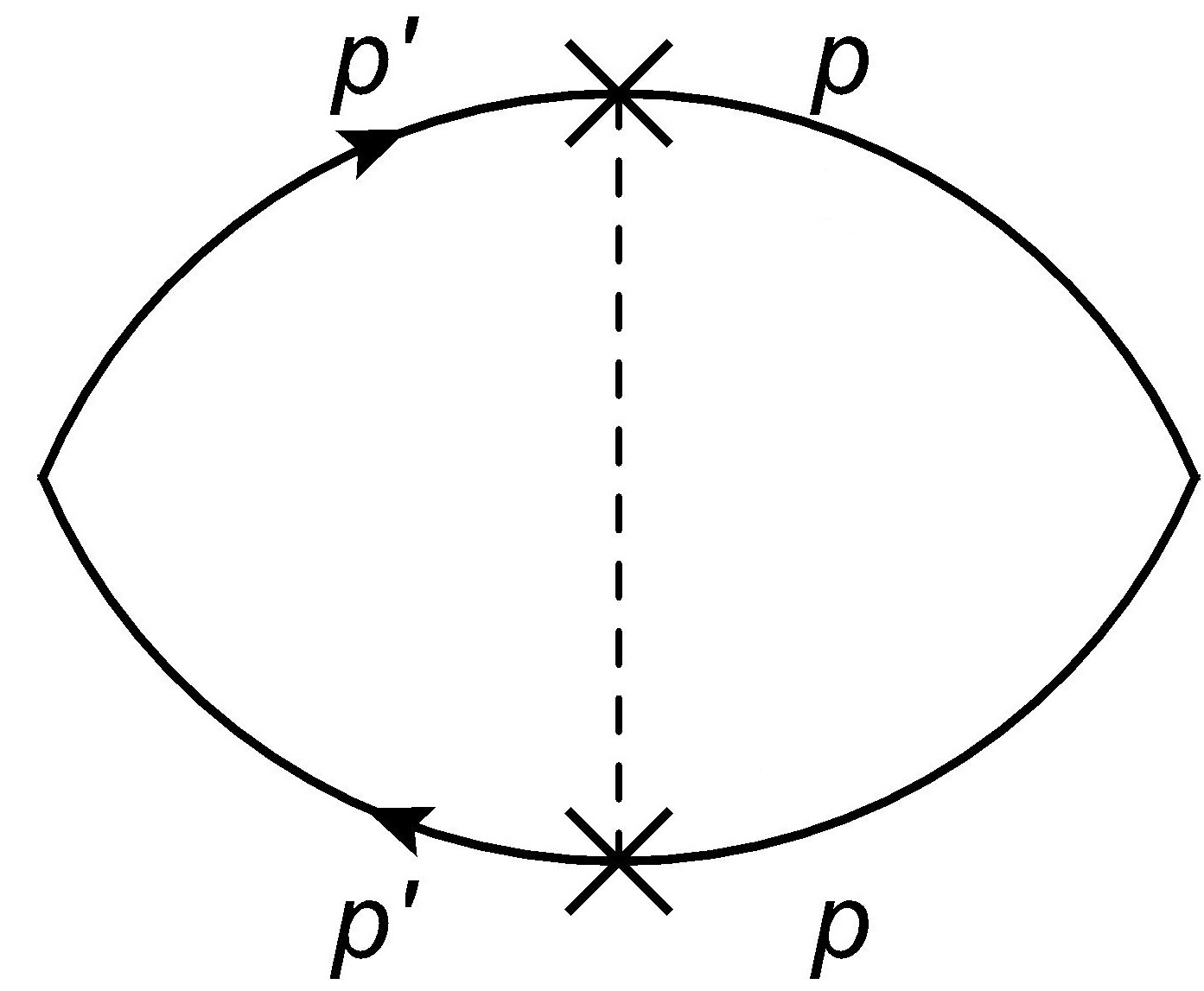}
		\caption{ The diagram needed to evaluate the extra vertex correction to the ISGE due to extrinsic SOC. The left and right vertices denote the spin vertex $S^y$ and the component $(p_x/m)$ of the number current vertex $J_x$, whereas, the crosses on the top and bottom Green functions line stand for $-i (\lambda_0^2/4) \pv' \times \pv$ and 
		 $-i (\lambda_0^2/4) \pv \times \pv'$, respectively.
		}
		\label{diagram_vertex}
	\end{center}
\end{figure}
with $S_{\alpha}^x=-e N_0 \alpha \tau E_x$. This is not the full story yet as we are to explain. What we have learned up to now is that the momentum dependence of the EY self-energy on the two spin-split Fermi surfaces yields an extra term to the Edelstein polarization. Such a momentum dependence can also modify the vertex corrections (the integrals $J_{\mu\upsilon}$ in Eq.(\ref{integrals})),  which lead to the renormalized spin vertex. To appreciate this aspect we notice that in evaluating such integrals in the absence of the RSOC, the moduli of ${\bf p}$ and ${\bf p'}$ are taken at the {\it unsplit} Fermi surface. We emphasize that, instead, taking into account the momentum dependence on the Rashba-split Fermi surfaces one gets an extra contribution.  Consider the diagram of Fig. \ref{diagram_vertex}. After integration over $\pv'$,  the left side part of the diagram gives
$$
-\frac{(\lambda_0^2/4)^2 p_F^2 p^2}{2\tau_0} \tau=-\frac{\tau}{2\tau_{EY}}\frac{p^2}{p_F^2}.
$$
If we  set $p=p_F$, we would recover  the standard diagrammatic calculation in the absence of intrinsic RSOC.
By combining the above left side with the rest of the diagram, one gets an additional contribution to the bare conductivity
\begin{eqnarray}
	(\delta \Pi)&=&-\frac{\tau}{2\tau_{EY}}
	\left(-\frac{e}{2\pi}\sum_{\bf p} \frac{ p^2}{p_F^2} {\rm Tr}\left[ G^A(\epsilon+\omega) \frac{\sigma^y}{2}G^R(\epsilon) \frac{p_x}{m}\right] \right)
	\nonumber \\
	&=&
	\frac{-\tau}{2\tau_{EY}}
	(\frac{e}{4mp_F^2}) 
	\left( \frac{p^3_+N_+}{-i\omega+\frac{1}{\tau_+}}-\frac{p^3_-N_-}{-i\omega+\frac{1}{\tau_-}}\right). \label{extra bare}
\end{eqnarray}
To this expression we must subtract the one obtained by replacing $p=p_F$, which is already accounted for in the ladder summation. Hence the extra vertex part ($\delta \Pi$) modifies the spin polarization to give the second contribution
\begin{equation}
( S^y)^{(2)}=\frac{1}{\left({\frac{\tau}{\tau_s}-i\omega\tau}\right)}\left(\frac{1-4i\omega\tau}{1-6i\omega\tau}\right)S_{\alpha}^x \left(-\frac{\tau}{2\tau_{EY}}\right).
	\label{extra_vertex}
\end{equation}
Hence, by summing the above result to Eq.(\ref{ed_self}),  the total spin polarization   reads
\begin{equation}
	\label{final_ed}
S^y=\frac{1}{\left(\frac{1}{\tau_s}-i\omega\right)} \left(1+\frac{2i\omega\tau}{1-6i\omega\tau}\right) S_{\alpha}^x \left(\frac{1}{\tau_{\alpha}}-\frac{1}{\tau_{EY}}
	\right)\approx \frac{1}{\left(\frac{1}{\tau_s}-i\omega\right)}S_{\alpha}^x \left(\frac{1}{\tau_{\alpha}}-\frac{1}{\tau_{EY}}
	\right).
	\end{equation}

In the diffusive regime, terms in $\omega\tau$ in the second round brackets on the right hand side of Eq.(\ref{final_ed}) which are responsible for higher-order frequency dependence, can be neglected. In the zero-frequency limit, the Eq.(\ref{final_ed}) has two main contributions described by the two terms in the last round brackets. The first term is responsible for the Edelstein result \cite{EDELSTEIN1990233} due to the intrinsic SOC, whereas the second one, which arises to order $\lambda_0^4$, is an additional contribution to the spin polarization due to the extrinsic SOC. 
In the Rashba model without extrinsic SOC, only the first term  is present and, indeed, Eq.(\ref{final_ed}) reduces to it when $\lambda_0=\omega=0$. 
After Fourier transforming, the above equation can be written in the form of the Bloch equation
\begin{eqnarray}
\partial_tS^y=-\left(\frac{1}{\tau_{\alpha}}+\frac{1}{\tau_{EY}}\right)S^y+\left(\frac{1}{\tau_{\alpha}}-\frac{1}{\tau_{EY}}\right)S^x_\alpha.
\end{eqnarray}
The terms on the right hand side describe the various  torques controlling the spin dynamics. The first term, which includes DP and EY contributions, is the spin relaxation torques, wheres the second term represent the spin generation torques.
The above result coincides with that obtained  in Ref.\cite{gorini2017theory} by the SU(2) gauge theory formulation.
We have then succeeded in showing by diagrammatic methods the origin of the EY-induced spin torque discussed by Ref.\cite{gorini2017theory}.
In the next Section we will generalize this result to the case when both RSOC and DSOC are present.

\section{Inverse spin-galvanic effect in the Rashba-Dresslhaus model}\label{R/D_model}
 
As we have seen in the previous Section, the size and form of the ISGE is greatly modified by the presence of the EY spin relaxation due to the extrinsic SOC. 
To analyze this fact more generally we focus here on the  model with RSOC and DSOC as well as SOC from impurities. In order to evaluate Eq.(\ref{Kubo}) for the Edelstein conductivity, we need the renormalized spin vertex $\Upsilon_i$. 
For vanishing RSOC or DSOC, the renormalized spin vertices share the same matrix structure of the bare ones $\Upsilon_i\sim \sigma_i$. However,
when both RSOC and DSOC are explicitly taken into account, $\Upsilon_x$ and $\Upsilon_y$ are not only simply proportional to $\sigma_x$ and $\sigma_y$, but also acquire components on both $\sigma_x$ and $\sigma_y$.
By following the  procedure shown in Eq.(\ref{vertex}) and upon integration over momentum, the vertex equation for $\Upsilon_y$ reduces to 
\begin{equation}
	\label{vertex_y}
	\begin{pmatrix}
		1-I_{00}+J_{00} &   -2(I_{yx}-J_{yx})\\
		-2(I_{xy}-J_{xy})        &   1-I_{00}+J_{00}
	\end{pmatrix} 
	\begin{pmatrix}
		\Upsilon^y_y\\
		\Upsilon^x_y
	\end{pmatrix}=
	\begin{pmatrix}
		1\\
		0
	\end{pmatrix}
\end{equation}
while that  for$\Upsilon_x$  is
\begin{equation}
	\begin{pmatrix}
		1-I_{00}+J_{00} &   -2(I_{xy}-J_{xy})\\
		-2(I_{yx}-J_{yx})        &   1-I_{00}+J_{00}
	\end{pmatrix} 
	\begin{pmatrix}
		\Upsilon^y_x\\
		\Upsilon^x_x
	\end{pmatrix}=
	\begin{pmatrix}
		0\\
		1
	\end{pmatrix},
	\label{vertex_x}
\end{equation}
where
\begin{eqnarray}\label{off_dia_rate}
	1-I_{00}+J_{00}&\simeq & \left(\frac{-i\omega+\langle \frac{1}{\tau_{\gamma}}\rangle +\frac{1}{\tau_{EY}}}{1-4i\omega\tau}\right)\tau\\
	-2(I_{xy}-J_{xy})&\simeq &\left(\frac{1-i\omega\tau}{1-4i\omega\tau}\right)\left(1-
	\frac{\tau}{\tau_{EY}}\right)\frac{2\tau}{\tau_{\alpha \beta}},
	\nonumber
\end{eqnarray}
where $\langle \dots \rangle$ indicated the average over the momentum directions. The technical points of the calculation in Eq.(\ref{off_dia_rate}) are given in appendix \ref{app} at the end of the paper. 
In the diffusive regime, $\frac{1}{\tau_{\gamma}}=(2m\gamma)^2D$ and $\frac{1}{\tau_{\alpha \beta}}=(2m)^2\alpha\beta D$ are the Dyakonov-Perel (DP) relaxation rates due to the total intrinsic spin-orbit strength and the interplay of RSOC/DSOC, respectively. For vanishing DSOC, the Eq.(\ref{off_dia_rate}) reduce to the same expression  in Eq.(\ref{vertexa}) as expected in the Rashba model. However, with both RSOC and DSOC, spin relaxation is anisotropic and one needs to diagonalize the matrix in the left hand side of Eqs.(\ref{vertex_y}-\ref{vertex_x}). Such a matrix then identifies the spin eigenmodes. Having in mind to derive the Bloch equations governing to spin dynamics, we rewrite Eq.(\ref{LRT}) by using Eq.(\ref{gamma_x1}) 
\begin{equation}
	\begin{pmatrix}
		S^x\\
		S^y
	\end{pmatrix}=
	\begin{pmatrix}
		\Upsilon^x_x &   \Upsilon^y_x\\
		\Upsilon^x_y        &   \Upsilon^y_y
	\end{pmatrix} 
	\sum_j
	\begin{pmatrix}
		\Pi_{xj}\\
		\Pi_{yj}
	\end{pmatrix}E_j
	\label{s xy}
\end{equation}
where, by virtue of Eqs.(\ref{vertex_y}-\ref{vertex_x})
\begin{equation}
	\begin{pmatrix}
		\Upsilon^x_x &   \Upsilon^x_y\\
		\Upsilon^y_x        &   \Upsilon^y_y
	\end{pmatrix} ^{-1}=\frac{\tau}{1-4i\omega\tau}
	\begin{pmatrix}
		-i\omega +\langle \frac{1}{\tau_{\gamma}}\rangle +\frac{1}{\tau_{EY}} &   \frac{2}{\tau_{\alpha \beta}}(1-i\omega\tau)\\
		\frac{2}{\tau_{\alpha \beta}} (1-i\omega\tau)   &   -i\omega +\langle \frac{1}{\tau_{\gamma}}\rangle +\frac{1}{\tau_{EY}}
	\end{pmatrix}.
	\label{Vertexx}
\end{equation}
In the diffusive regime we can safely neglect the factor $\omega\tau$ with respect to unity in the denominator in front of the matrix and in the off diagonal elements of the matrix.
The quantities $\Pi_{\rho j}$ appearing in  the right hand side of Eq.(\ref{s xy})  can be evaluated by standard techniques.  However, some care is required when evaluating the momenta due to the extrinsic SOC at the spin-split Fermi surfaces, as we did in Eq.(\ref{extra bare}).
The final result for the bare conductivities reads
\begin{eqnarray}
	\Pi_{xx}&=&
	\frac{-\tau S_\beta^x}{1-6i\omega\tau}\langle\frac{1}{\tau_{\gamma}}-\frac{1}{\tau_{EY}}-\frac{2}{\tau_{\gamma}}\frac{\alpha^{2}}{\gamma^2}\rangle, \label{barea}
	\\
	\Pi_{xy}&=& 
	\frac{-\tau S_\alpha^y}{1-6i\omega\tau}\langle\frac{1}{\tau_{\gamma}}-\frac{1}{\tau_{EY}}-\frac{2}{\tau_{\gamma}}\frac{\beta^{2}}{\gamma^2}\rangle, \label{bareb}
	\\
	\Pi_{yx}&=&  
	\frac{\tau S_\alpha^x}{1-6i\omega\tau}\langle\frac{1}{\tau_{\gamma}}-\frac{1}{\tau_{EY}}-\frac{2}{\tau_{\gamma}}\frac{\beta^{2}}{\gamma^2}\rangle, \label{barec}
	\\
	\Pi_{yy}&=& 
	\frac{\tau S_\beta^y}{1-6i\omega\tau}\langle\frac{1}{\tau_{\gamma}}-\frac{1}{\tau_{EY}}-\frac{2}{\tau_{\gamma}}\frac{\alpha^{2}}{\gamma^2}\rangle, \label{bared}
\end{eqnarray}
 with
\begin{eqnarray}
	S_\beta^x&=&-eN_0\tau \beta E_x\label{Ed_bx}\\
	S_\alpha^y&=&-eN_0\tau\alpha E_y\label{Ed_ay}\\
	S_\alpha^x&=&-eN_0\tau\alpha E_x\label{Ed_ax}\\
	S_\beta^y&=&-eN_0\tau\beta E_y.\label{Ed_by}.
\end{eqnarray}

We take the angular average over the DP relaxation rates in Eqs.(\ref{Vertexx}-\ref{bared})
\begin{equation}
	\int\limits_{0}^{2\pi}\frac{d\phi}{2\pi}\frac{1}{\tau_{\gamma}}=\frac{1}{\tau_{\alpha}}+\frac{1}{\tau_{\beta}}
\end{equation}
\begin{equation}
	(-2)(\alpha^2 \ {\rm or} \  \beta^2)\int\limits_{0}^{2\pi}\frac{d\phi}{2\pi}\frac{1}{\tau_{\gamma}}\frac{1}{\gamma^2}= \frac{-2}{\tau_{\alpha}} \ {\rm or} \  \frac{-2}{\tau_{\beta}}.
\end{equation}
where $\frac{1}{\tau_{\alpha}}=(2m\alpha)^2D$, $\frac{1}{\tau_{\beta}}=(2m\beta)^2 D$  are the DP relaxation rates due to RSOC and DSOC in the diffusive approximation. By inserting the above expression into Eqs.(\ref{barea}-\ref{bared}) and vertex correction in Eq.(\ref{Vertexx}) and using  Eq.(\ref{s xy}), we 
may write the expression of the ISGE components in a form reminiscent of the Bloch equations 
\begin{equation}
   \begin{pmatrix}
		-i\omega+\frac{1}{\tau_{\alpha}}+\frac{1}{\tau_{\beta}}+\frac{1}{\tau_{EY}} &   \frac{2}{\tau_{\alpha \beta}}\\
		\frac{2}{\tau_{\alpha \beta}}    &   -i\omega+\frac{1}{\tau_{\alpha}}+\frac{1}{\tau_{\beta}}+\frac{1}{\tau_{EY}}
	\end{pmatrix}
	\begin{pmatrix}
		S^x\\
		S^y
	\end{pmatrix}=
	\begin{pmatrix}
		-S_\alpha^y(\frac{1}{\tau_{\alpha}}-\frac{1}{\tau_{\beta}}-\frac{1}{\tau_{EY}})-S_\beta^x(\frac{-1}{\tau_{\alpha}}+\frac{1}{\tau_{\beta}}-\frac{1}{\tau_{EY}})\\
		S_\alpha^x(\frac{1}{\tau_{\alpha}}-\frac{1}{\tau_{\beta}}-\frac{1}{\tau_{EY}})+S_\beta^y(\frac{-1}{\tau_{\alpha}}+\frac{1}{\tau_{\beta}}-\frac{1}{\tau_{EY}})
	\end{pmatrix}\label{Bloch eq},                     
\end{equation}

Indeed, by performing the anti-Fourier transform with respect to the frequency $\omega$, Eq.(\ref{Bloch eq}) can be written as 
\begin{eqnarray}\label{BlochF}
\partial_t{\bf S}=-(\hat{\Gamma}_{DP}+\hat{\Gamma}_{EY}){\bf S}+(\hat{\Gamma}_{DP}-\hat{\Gamma}_{EY})\frac{N_0}{2}{\bf B},
\end{eqnarray}
where ${\bf B}$ represents the internal SOC field induced by the electric current. The $\hat{\Gamma}_{DP}$ and $\hat{\Gamma}_{EY}$ are the DP and EY relaxation matrix
\begin{eqnarray}
{\bf B}=2e\tau\begin{pmatrix}
\beta E_x+\alpha E_y\\
-(\alpha E_x+\beta E_y)
\end{pmatrix},
\hat{\Gamma}_{DP}=\begin{pmatrix}
\frac{1}{\tau_{\alpha}}+\frac{1}{\tau_{\beta}} & \frac{2}{\tau_{\alpha \beta}}\\
 \frac{2}{\tau_{\alpha \beta}} & \frac{1}{\tau_{\alpha}}+\frac{1}{\tau_{\beta}}
\end{pmatrix},
\hat{\Gamma}_{EY}=\begin{pmatrix}
\frac{1}{\tau_{EY}} & 0\\
0 & \frac{1}{\tau_{EY}}
\end{pmatrix}.
\end{eqnarray} 
Eq.(\ref{BlochF}) is the main result of our paper. It shows that  the  intrinsic and extrinsic SOC act in parallel as far as relaxation to the equilibrium state is concerned,
i.e. the DP and EY spin relaxation matrices add up. However, as far as the spin generation torques are concerned, DP and EY processes have opposite sign. This is in full agreement with the result of  Ref.\cite{gorini2017theory} once we take into account also the spin generation torque due to side-jump and skew-scattering processes discussed diagramatically in Ref.\cite{maleki2016edelstein}.  This is simply obtained by multiplying the DP relaxation matrix $\hat\Gamma_{DP}$ in the second term
in the right hand side of Eq.(\ref{BlochF}) by the factor $1+\theta^{sH}_{ext}/\theta^{sH}_{int}$, where $\theta^{sH}_{ext}$ and $\theta^{sH}_{int}$ are the spin Hall angles for extrinsic and intrinsic SOC.

To develop some quick intuition, one may notice that again for $\beta=\lambda_0=0$ and $E_y=\omega=0$, Eq.(\ref{Bloch eq}) reproduces the Edelstein result for the Rashba model \cite{edelstein1990solid}. Furthermore, when also $\omega\neq0$ it reproduces the frequency-dependent spin polarization for the Rashba model as shown in the previous Section. When $\lambda_0 \neq 0$ and $\beta=0$, we see that the ISGE,
due to the interplay of the extrinsic and intrinsic SOC, gets an additional spin torque, suggesting that the EY spin-relaxation is detrimental to the Edelstein effect.
The diagrammatic analysis reported here provides the following interpretation. The EY spin relaxation depends on the Fermi momentum. When there are two Fermi surfaces with different Fermi momenta, the one with the smaller momentum undergoes less spin relaxation of the EY type than the one with larger momentum. On the other hand, the ISGE arises precisely because there is an unbalance among the two Fermi surfaces with respect to spin polarization. For a given momentum direction, the larger Fermi surface contributes more to the Edelstein polarization than the smaller Fermi surface. Hence, the combination of these two facts suggests a negative effect from the interplay of Edelstein effect and EY spin relaxation. By neglecting the EY relaxation, one sees that  the DP terms can cancel each other if the RSOC and DSOC strengths are equal.  This cancellation or anisotropy of the spin accumulation could be used to determine the absolute values of the RSOC and DSOC strengths under spatial combination of spin dependent relaxation. \\ 
Finally, we comment on the relevance of our theory with respect to
existing experiments  Ref.\cite{PhysRevLett.112.056601}.  The latter show that the current-induced spin polarization does not align along the internal magnetic field
${\bf B}$  due to the SOC. According to our Eq.(\ref{BlochF}) this may occur due to the presence of the extrinsic SOC both in the spin relaxation torque and in the spin generation torque. Indeed when the extrinsic SOC is absent, the spin polarization must necessarily align along the ${\bf B}$ field. Hence, our theory could, in principle, provide a method to measure the relative strength of intrinsic and extrinsic SOC.

\section{Conclusions}\label{conclusions}
In this present work, we showed how the interplay of  intrinsic and extrinsic spin-orbit coupling modifies the current-induced spin polarization in a 2DEG. This phenomenon, known as the inverse spin galvanic effect,  is the consequence of the coupling between  spin polarization and electric current, due to restricted symmetry conditions.
 We derived the frequency-dependent spin polarization response, which allowed us to obtain the Bloch equations 
 governing the spin dynamics of carriers.  We   identified the various sources of spin relaxation. In fact, the precise relation between the non-equilibrium spin polarization and spin-orbit coupling depends on ratio of the DP and EY spin relaxation rates. More precisely, the spin-orbit coupling affects the spin relaxation time by adding the EY mechanism to the DP and, furthermore, it changes the non-equilibrium value of the current-induced spin polarization by introducing an additional spin torque. Our treatment, which  is valid at  the level of Born approximation and was obtained by diagrammatic technique agrees with  the analysis of Ref.\cite{gorini2017theory}, derived via the quasiclassical Keldysh Green function technique. Finally, to make comparison between  theory and experiments, we found that the spin polarization and internal magnetic field will not be aligned if the EY is strong enough. 
  
\appendixsections{one}
\appendix
\section{Integrals of  products involving pairs of retarded and advanced Green functions}\label{app}
To perform the calculations of the renormalized spin vertex in Eq.(\ref{off_dia_rate}) and also in all the analysis, we encounter the following kind of  integrals, which are evaluated to  first order in $\frac{\gamma}{v_F}$ and $\omega\tau$
\begin{eqnarray}
\sum_{\pv} p^n G^R_{\pm}(\epsilon+\omega)G^A_{\pm}(\epsilon)& \approx & 2\pi N_{\pm} p^n_{\pm}\frac{1}{-i\omega+\frac{1}{\tau_{\pm}}}\\
\sum_{\pv}p^n G^R_{\pm}(\epsilon+\omega)G^A_{\mp}(\epsilon)&\approx&2\pi N_0 p^n_{\pm} \frac{1}{-i\omega\pm2 i \gamma p_F+\frac{1}{\tau}}
\end{eqnarray}
where $n=0,1$.  We can then evaluate the $I_{00}$ integral as 
\begin{eqnarray}
I_{00} &=&\frac{1}{2\pi N_0 \tau_0}\sum_{\pv'}G^A_0(\epsilon+\omega) G^R_0(\epsilon)\nonumber\\
&=& \frac{1}{2\pi N_0 \tau_0}\sum_{\pv'}\frac{1}{4}\left(G^A_+(\epsilon+\omega) G^R_+(\epsilon)+G^A_+(\epsilon+\omega) G^R_-(\epsilon)+G^A_-(\epsilon+\omega) G^R_+(\epsilon)+G^A_-(\epsilon+\omega) G^R_-(\epsilon)\right)\nonumber\\
&=&\frac{1}{4N_0\tau_0}\langle\frac{N_+}{-i\omega+\frac{1}{\tau_+}}+\frac{N_-}{-i\omega+\frac{1}{\tau_-}}+\frac{N_0}{-i\omega+2ip_F\gamma+\frac{1}{\tau}}+\frac{N_0}{-i\omega-2ip_F\gamma+\frac{1}{\tau}}\rangle\nonumber\\
&\approx&(\frac{\tau}{\tau_0})\left(\frac{1-3i\omega\tau-\langle\frac{\tau}{\tau_{\gamma}}\rangle}{1-4i\omega\tau}\right)
\end{eqnarray}
and the same calculations for $2I_{xy}=2 I_{yx}$ yields
\begin{eqnarray}
2I_{xy}&=& \frac{2}{2\pi N_0 \tau_0}\sum_{\pv'}G^A_x(\epsilon+\omega) G^R_y(\epsilon)\nonumber\\
&=& \frac{2}{2\pi N_0 \tau_0}\left(\frac{-\alpha\beta}{4\gamma^2}\right)\sum_{\pv'}\left(G^A_+(\epsilon+\omega) G^R_+(\epsilon)-G^A_+(\epsilon+\omega) G^R_-(\epsilon)-G^A_-(\epsilon+\omega) G^R_+(\epsilon)+G^A_-(\epsilon+\omega) G^R_-(\epsilon)\right)\nonumber\\
&\approx&(\frac{4\tau}{\tau_0})(\frac{2\tau}{\tau_{\gamma}})\left(\frac{-\alpha\beta}{4\gamma^2}\right)(\frac{1-i\omega\tau}{1-4i\omega\tau})\nonumber\\
&=&\frac{2\tau}{\tau_{\alpha\beta}}(\frac{1-i\omega\tau}{1-4i\omega\tau}).
\end{eqnarray}

\acknowledgments{We thank Cosimo Gorini, Ilya Tokatly, Ka Shen and Giovanni Vignale for discussions. A.M. thanks Juan Borge for help received during the initial stages of this work.}

\end{document}